\documentclass[smallextended]{svjour3}       
\smartqed  
\usepackage{graphicx}
 \usepackage{mathptmx}      
\usepackage{latexsym}
\usepackage{enumerate}
\usepackage{textcomp}

\begin{document}

\title{The ARIEL Instrument Control Unit design}
\subtitle{for the M4 Mission Selection Review of the ESA\textquotesingle s Cosmic Vision Program}



\author{M. Focardi \and E. Pace \and M. Farina \and \\A. M. Di Giorgio \and J. Colom\'e Ferrer \and I. Ribas  \and \\C. Sierra Roig \and L. Gesa Bote \and J. C. Morales \and \\J. Amiaux \and C. Cara \and J. L. Augures \and \\E. Pascale \and G. Morgante \and V. Da Deppo \and \\M. Pancrazzi \and V. Noce  \and S. Pezzuto \and \\M. Frericks \and F. Zwart
         \and  \\ G. Bishop \and K. Middleton \and P. Eccleston \and \\ G. Micela \and G. Tinetti 
}


\institute{\textbf{Mauro Focardi} \at
              INAF - Osservatorio Astrofisico di Arcetri - 50125 Firenze, $ITALIA$\\
              Largo E. Fermi 5 \\
              Tel.: +39-055-275 5213; 
              \email{\emph{mauro@arcetri.astro.it}}           
           \and
           \textbf{Emanuele Pace \and Maurizio Pancrazzi  \and Vladimiro Noce} \at
           Universit\`a degli Studi di Firenze - Dipartimento di Fisica e Astronomia - 50125 Firenze, $ITALIA$;  
           \and
           \textbf{Anna Maria Di Giorgio \and Maria Farina \and Stefano Pezzuto}\at
           INAF - Istituto di Astrofisica e Planetologia Spaziali - 00133 Roma, $ITALIA$;
           \and
           \textbf{Joseph Colom\'e Ferrer \and Ignasi Ribas \and Carles Sierra Roig \and Louis Gesa Bote \and 	Juan Carlos Morales}\at
           ICE - Institut de Ci\`encies de l'Espai - 08193 Barcelona, $ESPA\widetilde{N}A$;
           \and
           \textbf{Jerome Amiaux \and Christophe Cara \and Jean Louis Augures} \at
           CEA - Commissariat \`a l'Energie Atomique - 91191 Saclay, $FRANCE$;  
           \and
           \textbf{Martin Frericks \and Frans Zwart} \at
           SRON - Netherlands Institute for Space Research - 3584 CA Utrecht, $THE$ $NETHERLANDS$;  
           \and
           \textbf{Enzo Pascale} \at
           Universit\`a degli Studi di Roma $"La$ $Sapienza"$ - 00185 Roma, $ITALIA$; 
           \and
            \textbf{Gianluca Morgante} \at
           INAF - Istituto di Astrofisica Spaziale e Fisica Cosmica - 40129 Bologna, $ITALIA$;   
           \and
            \textbf{Vania Da Deppo} \at
           CNR IFN LUXOR - Istituto di Fotonica e Nanotecnologie, 35131 Padova, $ITALIA$;   
            \and
           \textbf{Georgia Bishop \and Kevin Middleton \and Paul Eccleston} \at
           RAL Space - Rutherford Appleton Laboratory - OX11 0QX Harwell Oxford, $UK$; 
            \and
      	 \textbf{Giuseppina Micela} \at
           INAF - Osservatorio Astronomico di Palermo - 90134 Palermo, $ITALIA$; 
           \and
            \textbf{Giovanna Tinetti} \at
           UCL - University College of London - WC1E 6BT London, $UK$.
            }

\date{Received: date / Accepted: date}

\maketitle

\begin{abstract}

The Atmospheric Remote-sensing Infrared Exoplanet Large-survey mission (ARIEL) \cite{Tinetti_1} is one of the three present candidates for the ESA M4 (the fourth medium mission) launch opportunity. The proposed Payload \cite{Eccleston_1}, \cite{Morgante_1}, \cite{Da_Deppo_0} will perform a large unbiased spectroscopic survey from space concerning the nature of exoplanets atmospheres and their interiors to determine the key factors affecting the formation and evolution of planetary systems.

ARIEL will observe a large number ($>$ 500) of warm and hot transiting gas giants, Neptunes and super-Earths around a wide range of host star types, targeting planets hotter than 600 K to take advantage of their well-mixed atmospheres. It will exploit primary and secondary transits spectroscopy in the $1.2-8$ $\mu m$ spectral range and broad-band photometry in the optical and Near IR (NIR).

The main instrument of the ARIEL Payload is the IR Spectrometer (AIRS) \cite{Amiaux_1} providing low-resolution spectroscopy in two IR channels: $Channel$ 0 ($CH_0$) for the $1.95-3.90$ $\mu m$ band and $Channel$ 1 ($CH_1$) for the $3.90-7.80$ $\mu m$ range. It is located at the intermediate focal plane of the telescope \cite{Da_Deppo_1}, \cite{Da_Deppo_2}, \cite{Da_Deppo_3} and common optical system and it hosts two IR sensors and two cold front-end electronics (CFEE) for detectors readout, a well defined process calibrated  for the selected target brightness and driven by the Payload\textquotesingle s Instrument Control Unit (ICU).

\end{abstract}

\section{Introduction}
\label{intro}

The ARIEL ICU design is conceived for scientific data pre-processing and to implement the commanding and control of the AIRS Spectrometer. The ICU is interfaced on one side with the instrument and on the other side (spacecraft, S/C, side) with both the Data Management System (DMS) and the Power Conditioning and Distribution Unit (PCDU), both belonging to the hosting platform.

The DMS is composed of the On-Board Computer (OBC) and Solid State Mass Memory (SSMM) operating as the main buffering memory for scientific data and HK telemetries before sending them to Ground, thanks to a communication system based on two $X-band$ transponders. For this reason, the ICU internal memories are basically conceived and designed for temporary local buffering and to support a reduced data handling as the AIRS scientific data, once properly pre-processed, are delivered to the SSMM.

This characteristic is exploited to simplify the unit electrical design, saving mass and power, for both the ICU architectures (baseline and alternative) designed at this stage to be interfaced respectively to US detectors or EU detectors by means of their customized CFEE, operating at cryogenic temperatures.

\begin{figure}[!h]
\begin{center}
 \includegraphics[width=12cm]{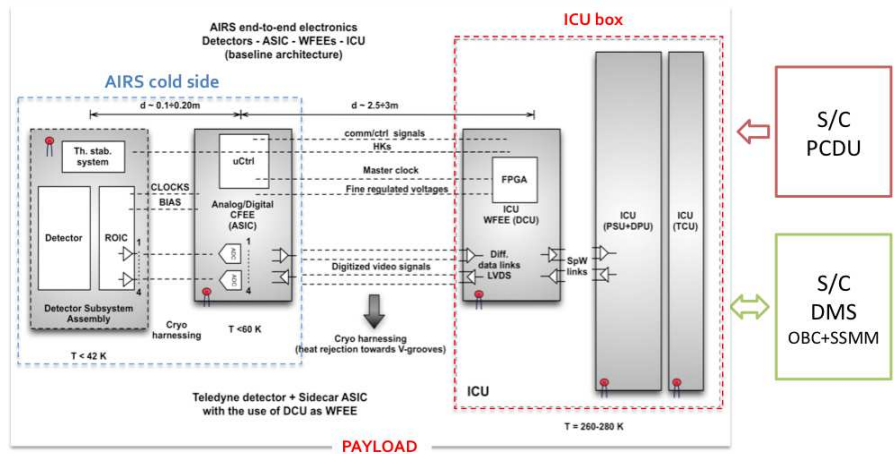}
\caption{AIRS detectors readout and digital processing electronics block diagrams.}
\label{fig:1}       
\end{center}
\end{figure}

As the ICU is a warm electronics, it will be located inside the S/C Service Vehicle Module (SVM) and connected to the AIRS CFEE by means of cryogenic harness.  The ICU subsystem I/F to the cryogenic harness is a warm FEE (WFEE), called Detector Control Unit (DCU), as shown in Fig. ~\ref{fig:1}.

In case of the adoption of US detectors from Teledyne (with a present higher TRL, Technology Readiness Level), the ARIEL CFEE will be represented by the SIDECAR\footnote{System Image, Digitizing, Enhancing, Controlling, And Retrieving.} ASIC, while in case of EU detectors, the CFEE will rely on a customized design presently under development by the \emph{SRON Space Research Institute} (The Netherlands).

The ICU responsibility and AIT/AIV activities at system level are in charge of Italy.

\section{AIRS detection modes}
\label{AIRS_det_mode}

\begin{figure}[!h]
\begin{center}
 \includegraphics[width=12cm]{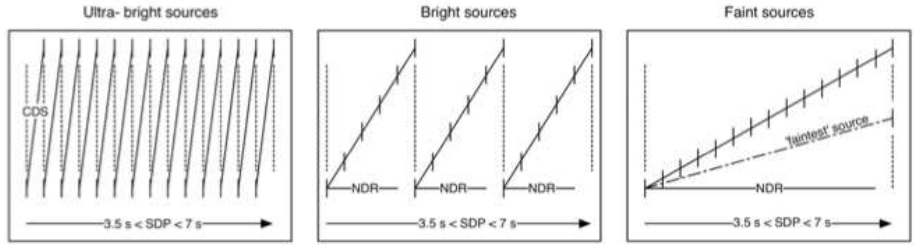}
\caption{The three different modes of detector read out depending on the target flux.}
\label{fig:2}       
\end{center}
\end{figure}

Due to the expected photon flux and its estimated dynamic range, different AIRS detection modes of operation will be implemented. The on-board data processing is in turn depending on the readout mode and thus three modes are defined: multiple CDS (Correlated Double Sampling) / Multiple slopes up-the-ramp sampling / Single slope up-the-ramp sampling. Fig. ~\ref{fig:2} illustrates these three modes of operation by representing the pixel-level signal.

The baseline selected detector (512 x 512 pixel with 15 to 18 $\mu m$ pixel pitch) for the ARIEL Spectrometer is similar to the Teledyne MCT (Mercury Cadmium Telluride) 1k x 1k array developed for the NASA\textquotesingle s NEOCam payload and based on the heritage of the WISE mission; this kind of detectors allow for non-destructive (or multi-accumulate sampling up-the-ramp) readout modes. This capability can effectively reduce the equivalent readout noise, improving the signal to noise ratio and allowing an easier identification and rejection of glitches in the signal induced by cosmic rays hits.

Scientific data, after the selected time sampling, are transferred from DCUs to DPU (Data Processing Unit). They are represented by images (detector windowing is foreseen) of 270 x 64 pixels for $CH_0$ and 100 x 64 pixels for $CH_1$ with 1 value (16 bit) per pixel and 1 Quality Criteria\footnote{The mean or $\chi ^2$ of the data is computed; deglitching by rejecting samples having a value above a threshold is an option (though not implemented currently, TBD with science SGS team).} (8 bits) per pixel, for a total of 24 bits for each computed ramp slope.

Assuming to sample up-the-ramp pixels in a non-destructive manner with the length of the ramp (duration) determined by the saturation limits of the detector, an estimation of the expected data rate can be provided, in principle, for any target of known flux (bright, medium, faint).

The overall daily data budget has been calculated and reported in Tab. ~\ref{tab:1}, where also the expected data rate for the housekeeping collected by ICU (in particular by the Telescope Control Unit, refer to the next paragraph) is provided within the following assumptions:

\begin{itemize}
\renewcommand{\labelitemi}{\tiny$\bullet$}
\item Fine Guidance System (FGS), VIS/NIRPhot and NIRSpec science channels \cite{Rataj_1} telemetries are not taken into account;
\item Different integration times could be used as a function of the targets brightness;
\item In case of very bright sources (e.g. 0.1 s of pixel saturation time), several exposures will be averaged together following a CDS readout scheme, to build one frame every $\sim$4 s.

\end{itemize}

The estimation includes the need of having a detector reset between each ramp.

The overall ARIEL daily data volume (25.0 Gibit/day, including FGS, NIRPhot and NIRSpec channels) is dominated by AIRS-$CH_0$, AIRS-$CH_1$ and NIR-Spec and takes into account the observing efficiency (95$\%$ - targets + calibration), as well as the required data volume margin (30$\%$ at this stage). 

The calculation provided in Tab. ~\ref{tab:1} assumes on-board fitting of the ramp, with an average ramp length to saturation of 3.26 s for the AIRS channels.

In particular, it is assumed that sampling up-the-ramp pixels is performed in a non-destructive manner with a relative high sampling rate ($\sim$3.5 Hz for AIRS-$CH_0$, $\sim$9.3 Hz for AIRS-$CH_1$), followed by destructive readouts after a well defined number of samples, depending on the brightness of the target.

Indeed, the actual read-out mode to be used will vary between targets (as a function of their brightness) with the possibility of setting the ramp integration time in the range of 3.5 to 7 s.

\begin{table}[!h]
\begin{center}
 \includegraphics[width=12cm]{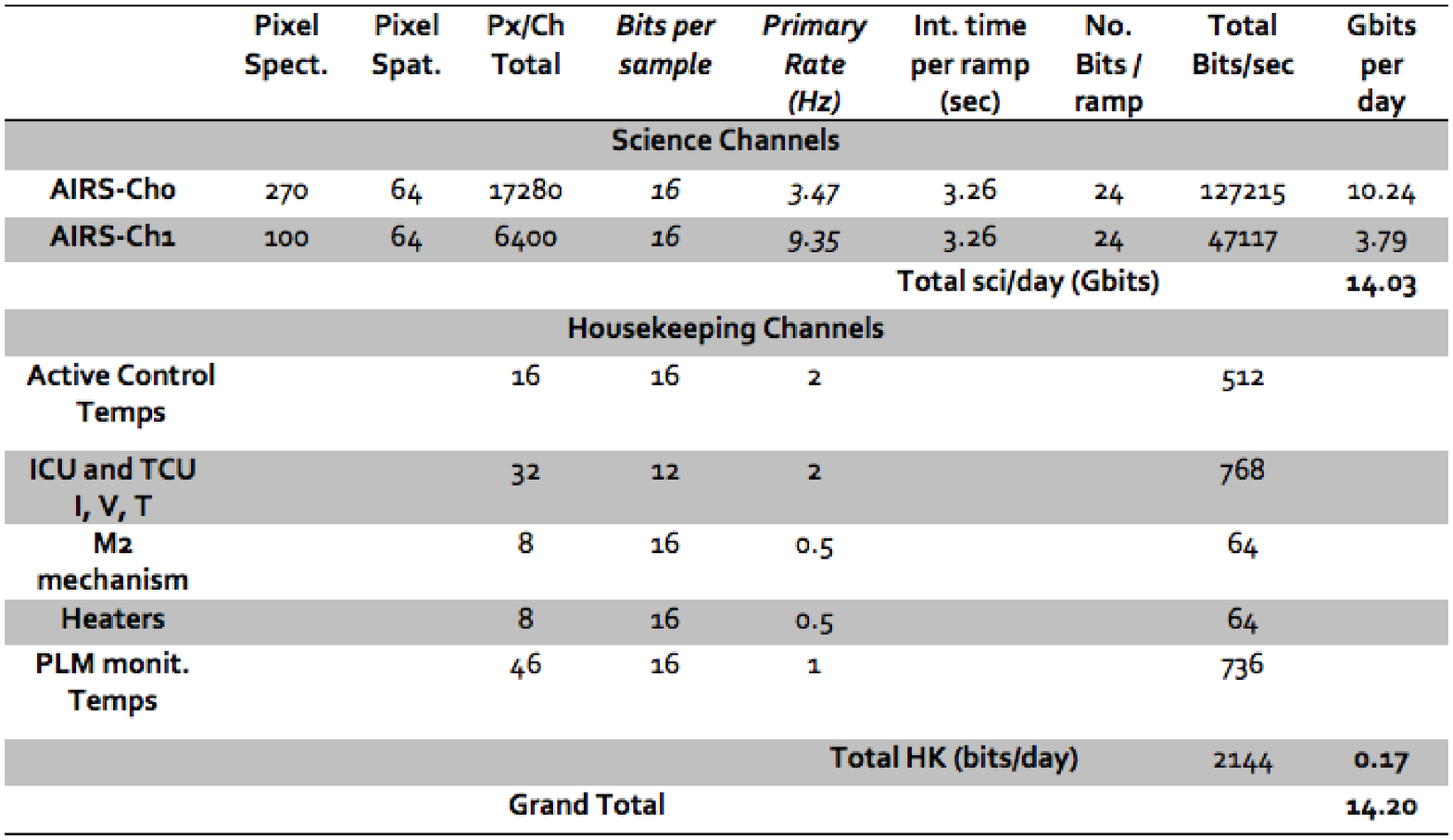}
\caption{Calculation of AIRS science and Payload HK data rate budgets.}
\label{tab:1}       
\end{center}
\end{table}

For each target in the ARIEL Targets List, the expected flux will be used for refining the best readout scheme to be driven by the ICU, obtaining the corresponding data rate.

The adopted scheduling tool for ARIEL observations, calibrations and data delivery to Ground, will also be used to show how the payload data rate may vary throughout the mission and to evaluate the expected maximum and average data rates, thus allowing for a correct dimensioning of the on-board processing (along with the ICU buffering capabilities) in quasi-real-time, prior to send data to the S/C SSMM.

\section{ICU baseline architecture}
\label{ICU_baseline}
\begin{figure}[!h]
\begin{center}
 \includegraphics[width=12cm]{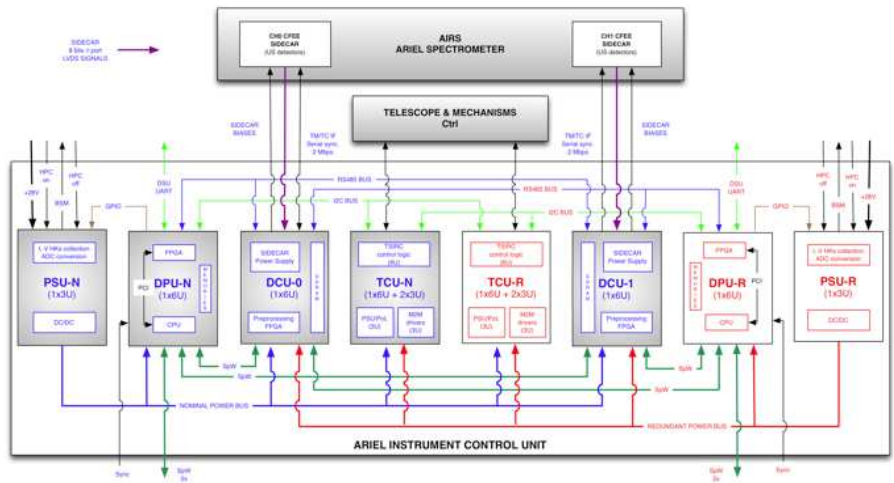}
\caption{ARIEL ICU and TCU baseline solution block diagram.}
\label{fig:3}       
\end{center}
\end{figure}

The ICU baseline architecture includes five (active or switched-on at the same time) units:

\begin{itemize}
\renewcommand{\labelitemi}{\tiny$\bullet$}
\item 1 PSU - Power Supply Unit
\item 1 DPU - Data Processing Unit
\item 2 DCU - Detector Control Unit
\item 1 TCU - Telescope Control Unit

\end{itemize}

as represented in Fig. ~\ref{fig:3} along with the number and type of needed PCBs (exploiting standard 3U and 6U formats).

The Telescope Control Unit is considered an ICU slave subsystem and for its complexity and required volume is located in an independent box, with stacked drawers to the unit main box. This configuration is exploited in both ICU baseline and alternative designs, as it presents several advantages.

Indeed, as currently foreseen, the TCU will be provided by Spain and shall host the main logic board called Thermal Stabilizer \& IR Calibrator (TSIRC), the M2 mirror mechanism (M2M) drivers and the needed power section and points of load (PoL) to properly feed its subsystems.
In particular, the Telescope Control Unit shall be able to accomplish the following tasks in ARIEL Science and Calibration modes:

\begin{itemize}
\renewcommand{\labelitemi}{\tiny$\bullet$}
\item driving the M2 refocusing mechanism 
\item driving the on-board calibration source
\item	monitoring the thermal state of several PLM elements 
\item	controlling the thermal stability of the Thermal Control System (TCS) for the following PLM subsystems:

\subitem	- AIRS detectors (actively cooled down to the operating temperature)
\subitem	- FGS, VIS/NIRPhot and NIRSpec detectors 
\subitem	- M1 mirror

\end{itemize}

The Telescope Control Unit will be composed of three separated boards, as the volume of the electronics to fulfil the Unit requirements is larger than a standard PCB and also to ease AIT/AIV activities at system level since the boards can operate independently with only a power supply (different for each unit) and their respective data buses.

In Fig. ~\ref{fig:3}, the ICU nominal units (TCU included) are indicated with blue labels, whilst the three redundant units are highlighted by means of red labels. They are hosted inside two stacked and independent boxes including PSUs, DPUs and DCUs (ICU box) and TSIRC logic, M2M drivers and the needed additional PSU (TCU box). The two boxes are electrically connected (power and TM/TC) by means of external harnessing exploiting front panel connectors to facilitate AIV/AIT activities, as the Units will be integrated and tested separately before integration. 

Both ICU and TCU boxes will implement their own back panel for routing power and signals lines connecting the internal electronics boards or, alternatively, will exploit external connections, but the latter solution would limit the allocated volume for the two boxes. A final assessment on both solutions will be performed during the next phase, taking into account the needed resources in terms of mass, volume, power dissipation and overall complexity. At the present time, in order to minimize the length and the mass of the harnessing connecting the cases, the stacked configuration is preferred.

The ICU baseline electrical architecture relies on the adoption of US detectors (H1RG-type) and cold front-end electronics (CFEEs) from Teledyne (SIDECAR), given their very high TRL and space heritage with respect to the present European alternative. The SIDECAR solution is the best one to drive properly the US MCT (HgCdTe) detectors and to save mass, volume and power at the same time. They can work easily down to the ARIEL required cryogenic temperatures ($\leq$ 60 K for SIDECARs and $\leq$ 42 K for detectors) so that both $CH_0$ and $CH_1$ are fed and controlled thanks to the adoption of two DCU boards, residing in the warm part of the Service Module. The two electronics sides will be connected by means of cryogenic harnessing, passing through the three $V$-grooves \cite{Morgante_1} (working at different temperatures) of the telescope assembly.

The present ICU architecture exploits a partial cold redundancy and cross-strapping capability. In particular, both TCUs and DCUs are cross strapped and can work along with PSU and DPU boards (Nominal and Redundant) as a whole, although DCUs aren\textquotesingle t involved in a cold redundant configuration as no duplicated DCUs are foreseen. A very similar ICU architecture, involving DCUs as SIDECAR I/F (for biases, clocks and control signals), has been already designed and adopted for the Euclid Mission (NISP Instrument). Each DCU controls and interfaces a single SIDECAR (as well as the related detector) and, in this sense, its design can be considered for the ARIEL Payload a strong heritage from the Euclid project. 

Indeed, an overall DPU/DCU/SIDECAR/H2RG detector chain reliability figure, higher than 98$\%$, has been computed and for this reason redundancy for ARIEL DCUs has not been considered, also because the related increasing complexity and needed budgets in terms of power, mass and volume. 

At the present time the DCU Technology Readiness Level (TRL) has been demonstrated higher than 5 and a DCU EM has been already manufactured and fully tested as it is working properly along with the SIDECAR and the detector. A DCU EQM model for NISP (very similar to the EM one) is under manufacturing and testing by the Italian industry. Moreover, a DCU/SIDECAR I/F simulator has been developed for the NISP instrument. The same philosophy concerning the DCU simulator is adopted for the ARIEL ICU case. 

As baseline, all the ICU and TCU boards (N and R) are designed respecting both the 3U (160 mm x 100 mm) and 6U (233 mm x 160 mm) standard PCB formats. The ICU boards will be stiffened by a proper mechanical frame with the external I/O connectors fixed and screwed to the board external panels. The TCU logic sections will be internally interfaced with a board implementing the M2M drivers and a board hosting the power supply and points of load required to feed them.

\begin{table}[!h]
\begin{center}
 \includegraphics[width=12cm]{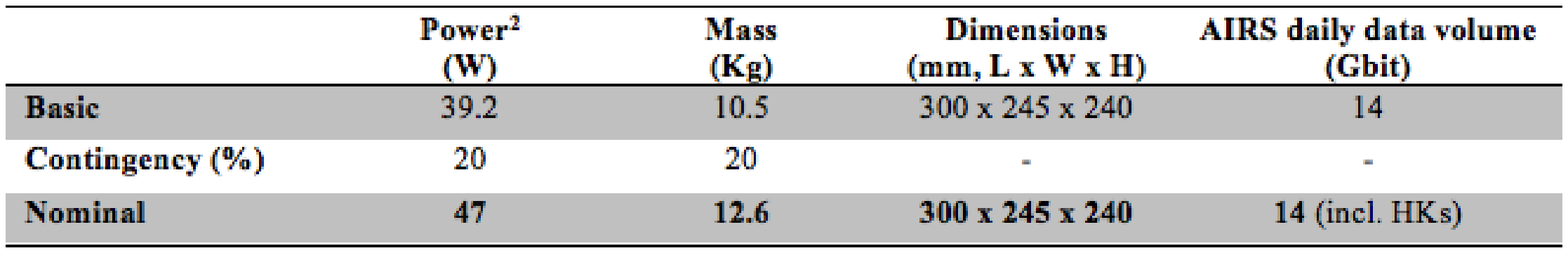}
\caption{ICU allocated budgets by ESA.}
\label{tab:2}       
\end{center}
\end{table}

\begin{table}[!h]
\begin{center}
 \includegraphics[width=12cm]{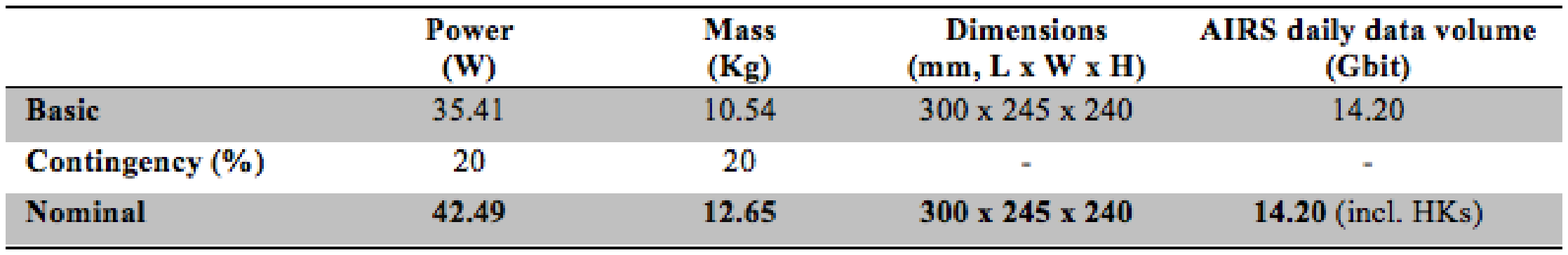}
\caption{ICU expected budgets, computed by the ARIEL Consortium.}
\label{tab:3}       
\end{center}
\end{table}

In particular, the ICU box shall host:

\begin{itemize}
\renewcommand{\labelitemi}{\tiny$\bullet$}
\item	two (N and R) 3U-format PSU boards, providing +5 V to ICU's DPU and DCU boards and +28 V filtered (N and R lines) to TCU box; 
\item	two (N and R) 6U-format DPU boards; 
\item	two ($CH_0$ and $CH_1$) not-redundant 6U-format DCU boards; 

\end{itemize}

and the TCU box:

\begin{itemize}
\renewcommand{\labelitemi}{\tiny$\bullet$}
\item	two (N and R) 3U-format PSU boards, locally deriving (thanks to on-board DC/DCs and PoL) $\pm$5 V and the needed voltage levels (+20 V, $\pm$12 V) from +28 V filtered coming from ICU; 
\item	two (N and R) 6U-format TSIRC boards (hosting a control FPGA, IR calibration source drivers, etc.); 
\item	two (N and R) 3U-format M2M driver boards (or a single 6U board hosting N and R drivers); 

\end{itemize}

for a total of nine equivalent (from the point of view of the overall dimensions and volume allocation) 6U-format boards, fitting the ESA allocated budgets (power\footnote{TCU included, Decontamination Mode excluded.}, mass and volume - refer to Tab. ~\ref{tab:2}  and ~\ref{tab:3}).
The lateral sides of the PCB modules will be equipped with card-lock retainers, used to fix them to the unit internal frame. All the boxes panels will be manufactured in an Aluminum alloy and then externally painted in black (except the bottom panel) to improve radiating exchange with the environment and assure, at the same time, a proper thermal conduction towards the SVM mounting panel.

\subsection{PSU board design}
\label{PSU}
The PSU board is a standard Power Supply Unit hosting DC/DC converters with a number of secondary sections needed to support the adopted cross-strapped and partially redundant configuration. It is in charge of collecting currents, voltages on secondary outputs and temperatures HK (A/D converted internally to the Unit, exploiting the SPI HK I/F for signals and control lines to/from the ADCs). The Unit consumption monitoring is in charge of platform as well as its switching on/off (both PSU and DPU boards together, thanks to a sequencing logic belonging to PSU), by means of HPC commands.

\begin{figure}[!h]
\begin{center}
 \includegraphics[width=12cm]{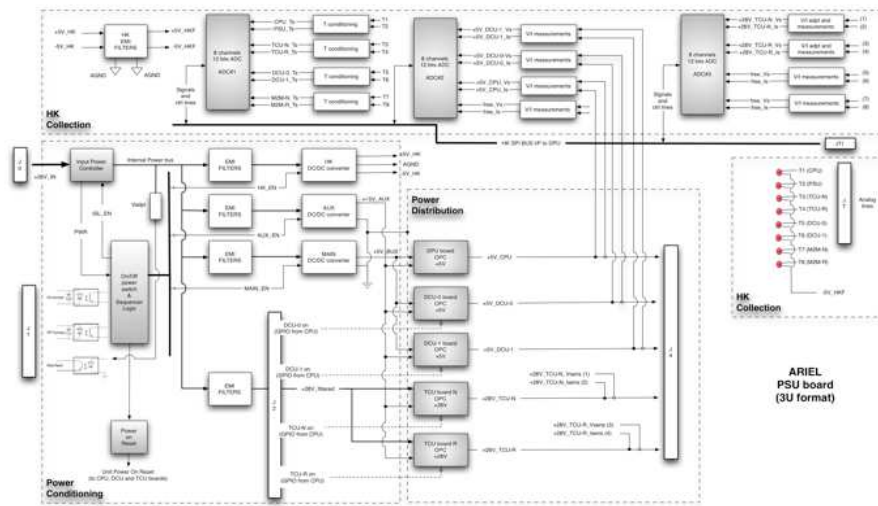}
\caption{Power Supply Unit block diagram.}
\label{fig:4}       
\end{center}
\end{figure}

The PSU is mainly composed of three sections (refer Fig. ~\ref{fig:4}):

\begin{enumerate}
\item Power conditioning section, performing the following tasks:

\subitem{

\begin{itemize}
\renewcommand{\labelitemi}{\tiny$\bullet$}
\item DC/DC conversion, i.e. main DC/DC for the generation of the +5 V to be distributed to the other boards, 			Aux DC/DC for internal logic powering, HK DC/DC for powering the HK section for the acquisition of 			voltages / currents / temperatures HK;
\item Inrush current limitation;
\item Polarity inversion protection;
\item Power-on sequence generation;
\item Unit power-on reset generation;
\item EMI (Electro-Magnetic Interference) filtering.

\end{itemize}
}

It is worth noting that only a main DC/DC converter is foreseen to feed respectively CPU (N or R) and DCU$_0$ + DCU$_1$ boards as presently it is assumed that it will comply with the overall required current. As alternative, a further DC/DC converter can be exploited to feed some of the itemized boards, provided that both DC/DC can satisfy the required current absorption and be accommodated on a 3U board at the same time.

\item Power distribution section hosting Output Power Controllers (OPC), implementing switching capabilities and overcurrent plus overvoltage protections on the +5 V and +28 V voltage/current distribution lines (+28 V only to the Telescope Control Units, N and R);

\item HK acquisition section with three 12 bits ADCs for voltages, currents and temperatures measurements, controlled and acquired by the processor via SPI (Serial Peripheral I/F).

\end{enumerate}

Each electronic board, apart TCU, is basically supplied by a main voltage level of +5 V protected for overvoltage and overcurrent and locally on-board (DPU and DCU) are derived, by means of a Point of Load (PoL), the secondary voltage levels needed by the hosted electronic components.

\subsection{DPU board design}
\label{DPU}

The Data Processing Unit can be implemented as a single 6U board hosting a CPU (the UT699E processor from Cobham, as baseline) and a co-processing FPGA, hosting some peripherals.

\begin{figure}[!h]
\begin{center}
 \includegraphics[width=12cm]{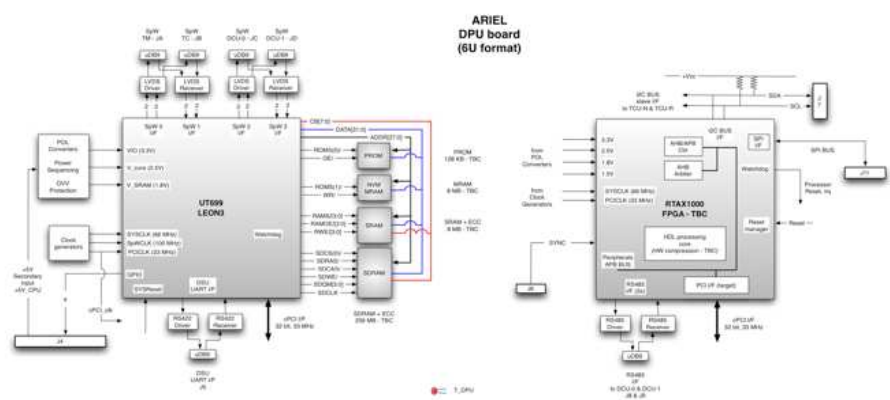}
\caption{Data Processing Unit block diagram.}
\label{fig:5}       
\end{center}
\end{figure}

Memories for:

\begin{enumerate}[i]
\item booting (PROM)
\item	storing the ASW (E2PROM and/or NVM e.g. MRAM)
\item	data buffering (e.g. SDRAM)
\item	data processing support (e.g. SRAM, SDRAM)

\end{enumerate}

are included in the design as well. The DPU board block diagram is provided in Fig. ~\ref{fig:5}. The two main blocks, i.e. the UT699E CPU and the RTAX1000 FPGA are connected through an on-board cPCI bus.

The selected enhanced UT699E CPU is a 32 bits fault tolerant LEON3FT SPARC V8 microprocessor supporting up to 100 MHz clock rate and allowing up to 140 DMIPS. The processor includes an on-chip Integer Unit (IU), a Floating Point Unit (FPU), a Memory Controller with a DMA Arbiter and a UART-based DSU I/F. It is interfaced to the on-board FPGA by means of a 32 bits wide, 33 MHz, cPCI bus supporting DMA (in case the SDRAM controller function were assigned to the FPGA).

One of the main characteristics of the adopted UT699E CPU is the on-board availability of 4 embedded SpaceWire (SpW) links (2 supporting the RMAP protocol) allowing to be directly interfaced to the SVM (OBC and SSMM Units) and to the DCU SpW I/F. The two SpW links implementing the RMAP protocol could be exploited to read from and write to the DCU FPGA registers\footnote{Indeed, the DCU SpW I/F link could be replaced with a serial I/F having reduced performance from the point of view of data rate, but the former offer the possibility to the CPU to read from and write to directly to the DCU FPGA registers, for remote configuration, thanks to RMAP protocol.}.

The DPU FPGA, along with the processor, is in charge of the DCU board management (by means of the RS485\footnote{The UT699E LEON3FT UART port is not compatible with the UART/RS485 hosted by the DCU FPGA as it isn\textquotesingle t able to manage the enabling/disabling of the transmission driver as required by the standard RS485 when adopting a single TX/RX line. For this reason, the RS485 bus, is used to interface the DCU FPGA by means of the Data Processing Unit FPGA.} bus I/F, as adopted for the Euclid\textquotesingle s DCU) and of the data acquisition (through the SpW I/F) and pre-processing tasks, e.g. implementation of the logic for the \emph{summing up the ramp} readout mode, data deglitching and lossless SW compression, if needed. Alternatively, some of the pre-processing tasks could be devolved to the SIDECAR ASIC or to the DCU unit in order to share properly the overall data processing load and the needed resources.
Analogously, to increase the overall ICU performances, the possibility to adopt an FPGA-based (or HW) implementation of the data compressor (and/or other processing routines) is under evaluation.

The DPU FPGA is interfaced to the TCU TSIRC board by means of an I$^2C$ bus I/F (for parameters configuration, telescope mirrors temperatures and mechanisms HK telemetries acquisition). An embedded HDL\footnote{Hardware Description Language.}-based Finite State Machine (FSM), in charge of controlling and scheduling the FPGA tasks, is foreseen along with an AMBA-bridged (AHB/APB) bus to connect and control all the internal peripherals thanks to an AHB arbiter.

Finally, an on-board PoL is included in the DPU design, with the aim of providing all the needed fine-regulated voltage levels to feed properly the processor and the FPGA (core voltages).

As alternative to the adoption of the UT699E as main processor, the Cobham/GR UT700 or the GR712RC dual-core LEON3FT CPU could be selected. The latter results to be one of the eligible on-board CPU for implementing both instrument control and data acquisition and processing functionalities (e.g. for SW data compression) exploiting properly its dual-core nature.

In particular, the GR712RC processor, supported by the RTEMS OS, can be more easily exploited in the so-called AMP (Asymmetric Multi-Processing mode) configuration (instead of SMP or Symmetric Multi-Processing mode), as the SW tasks can run asynchronously on the individual cores, configured to have separate memory addressable areas and hardware resources. 

This choice is normally driven by the fact that in space applications a high level of reliability and testability is needed, with a deterministic behaviour of the SW, and the latter can be better achieved by means of the AMP mode, where any hypothetical anomaly or overload of the running tasks in the additional core would not affect the effective reactivity of the other one, granting a physical insulation of the running spaces.

In case of AMP operating mode, the multi-core design requires extra-work to manage the possibility of concurrent access to all the shared resources (interrupts, timers, peripherals, memory). In order to use a multi-core processor, the software should be split up and distinguished into items that can run in parallel on the different cores.

In this configuration, two instances of the RTEMS OS in AMP mode are executed. The RTEMS running on the first GR712RC core, the boot processor, has the control over the primary resources and initializes the overall environment, while the RTEMS running on the second core has not access to the main resources but keep a full and independent control over its own threads scheduling, being the management of the other resources left to the developer choices.

It should be noted that the GR712RC can exploit up to 6 embedded SpW I/F if no SDRAM-type memory is directly interfaced, otherwise only 4 links are available, as in case of the UT699E processor.

The final and proper selection of the processor to be adopted for the management of the AIRS Spectrometer will be performed during the next phase of the Project, when the overall requirements on instrument management and data processing as well shall be finely addressed. Indeed, the choice of the AMP dual-core architecture should be justified by the actual need in terms of CPU resources (mainly peripherals, as the GR712RC too guarantees up to 140 DMIPS, when running at 100 MHz).

\subsubsection{DPU SW}
\label{DPU_SW}
As described above, the DPU science data handling functionalities include the AIRS spectrometer digital data (16 bits/pixel, 24 bits depth for ramps\footnote{Some more bits, beside 16, are needed to represent the quality criteria of the ramps slopes and fitting. AIRS adopts additional 8 bits for the ramp quality criteria definition.}) acquisition, buffering and pre-processing. A task of lossless compression (e.g. adopting the RICE algorithm, providing a compression ratio CR=2 at least) could be planned as well, although it is not strictly required. The compressed data are packetized according to the CCSDS protocol format and sent to the S/C DMS for storing and later downloading to Ground. Pre-processing and compression tasks can be disabled in case of raw data request from the Spacecraft/Ground (ESA mandatory requirement).

The science data handling functionalities will be implemented on the ICU\textquotesingle s Application Software (ASW), running on the DPU CPU. It handles all the ICU / Spectrometer and ICU / TCU digital interfaces and implements the following instrument monitoring and control functionalities: verifying and executing the telecommands received from the S/C, handling the switching on/off of the ICU and TCU subsystems, configuring and commanding the spectrometer sub-units, monitoring the ICU and AIRS units, reporting housekeeping and events, supporting the payload FDIR (Fault Detection, Isolation and Recovery) tasks and the operational modes, managing the on-board time thanks to a combination of the absolute time (received from the S/C through the SpaceWire protocol and Time Codes\footnote{It is also foreseen, as baseline, an external sync signal (with TBD frequency, amplitude and overall characteristics) in case the SpW packets time stamping exploiting Time Codes and an internal HW clock were not able to guarantee the needed timing accuracy for scientific data processed and sent to the S/C in quasi real-time.}) and the internal time (based on a HW clock).

The listed functionalities will be implemented by means of the CCSDS PUS services and all the mandatory PUS services will be guaranteed along with a set of services specific to the ARIEL Mission (private services for the ARIEL electronics subsystems).

\subsection{DCU board design}
\label{DCU}
\begin{figure}[!h]
\begin{center}
 \includegraphics[width=12cm]{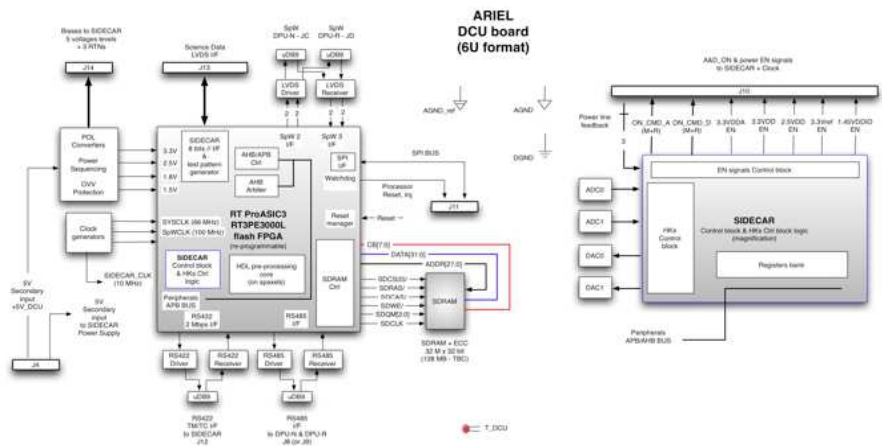}
\caption{Detector Control Unit block diagram.}
\label{fig:6}       
\end{center}
\end{figure}

The proposed Detector Control Unit design (refer to Fig. ~\ref{fig:6}) is a heritage of the design adopted for the DCUs of the NISP instrument on-board the Euclid Mission, where the same kind of detectors have been used along with the same CFEEs (SIDECARs). This choice allows minimizing all the risks concerning the design, development, performance characterization and testing activities on the board. The DCU hosts, as baseline, a FLASH-based reprogrammable FPGA to offer maximum flexibility also in case of late requirements specification (or modification) from the ARIEL Science Team. Alternatively, a Microsemi RTAX-family FPGA (in anti-fuse technology and so not reprogrammable once burned, as One Time Programmable -OTP- logic) could be adopted in case of early requirements specification and detectors / CFEEs selection.

The FPGA presently selected is a Microsemi ProASIC3-type device offering the capability to embed a HDL FSM with some programmable Science data pre-processing tasks (e.g. pixels co-adding, ramp slopes computing etc.) by means of a flexible parameters configuration that can be reprogrammed up to the EQM/FM unit. The FPGA also hosts a SDRAM memory controller to manage 128 MB of on-board memory used as a buffer to support the HDL-based pre-processing tasks.

It should be noted, indeed, that for the Euclid Mission case the use of a reprogrammable FPGA for the logic device implementing the interface with the detectors system has been preferred for the following reasons:

\begin{itemize}
\renewcommand{\labelitemi}{\tiny$\bullet$}
\item	Request of maximum flexibility from the Science Team along with the DCU development process; 
\item	Lack of knowledge of the actual behaviour of the logic interface to the SIDECAR: unexpected behavior could also have required mitigation in the FPGA not known a-priori during the following assessment phase;
\item	Risk of late modifications required on the FPGA design (e.g. concerning the implementation of updated high-performance pre-processing tasks): the chosen RTProASIC3 FPGA allows for the modification of the design via a JTAG port without opening the unit box and changing/removing the device.

\end{itemize}

On the other hand, the use of a flash reprogrammable FPGA has the disadvantage of a lower level of immunity to radiation effects with respect to a FPGA based on anti-fuse technology (e.g. Microsemi RTAX-S family) and its radiation-hard design improvement requires a not negligible effort (e.g. I/O, at RTL level, logic placement, etc.), although for the L2 radiation environment 50 krad outside the S/C can be assumed as a typical value.

For this reason, the standard rad-hard FPGA design flow should be modified with the introduction of radiation mitigation activities up to the validation with EM and EQM models, but this risk can be properly assessed and addressed by the ARIEL Consortium, as at least a European company has already acquired all the needed knowledge and competences to interface and drive the Teledyne SIDECAR + H1RG detector system.

The DCU WFEE is in charge of SIDECAR clocking (at least a master clock is needed for the ASIC) and feeding (secondary finely regulated voltages produced by an on-board PoL, refer to Tab. ~\ref{tab:4}) and it collects digitised scientific data and HK (currents, voltages and temperatures) describing the ASIC status. The needed enabling and control signals for SIDECAR management are represented in Fig. ~\ref{fig:6}, on the magnification of the box inside the FPGA block diagram (on the right). Three different grounding references (analog and digital) are foreseen for a clean power supply feeding.

In particular, the SIDECAR Science I/F is based on an 8 bits LVDS parallel I/F (with data buffering and packets CRC) and a TM/TC I/F running @ 2 Mbps (serial syncro) + master clock line @ 10 MHz. The DPU I/F shall be based, instead, on SpW for Science data TM along with a RS485 serial I/F offering the capability to manage and configure the DCU FPGA from DPU. Alternatively, the FPGA registers could be managed thanks to the RMAP protocol exploiting the SpW-based I/F.

An important issue of the electrical I/F to the SIDECAR ASICs is the harness, electrically and thermally linking the WFEE part of the electronics (working at the Service Module temperature of 270-300 K) and the CFEE part (working at T $<$ 60 K, on the PayloadÕs optical bench). For this kind of harnessing it is foreseen to split the electrical connections in different parts, or mated cables, characterized by different thermal conductivities (e.g. copper, constantan or manganine, phosphor bronze, steal, etc.) in order to be properly connected to the three $V$-grooves heat sinks.

\begin{table}[!h]
\begin{center}
 \includegraphics[width=11cm]{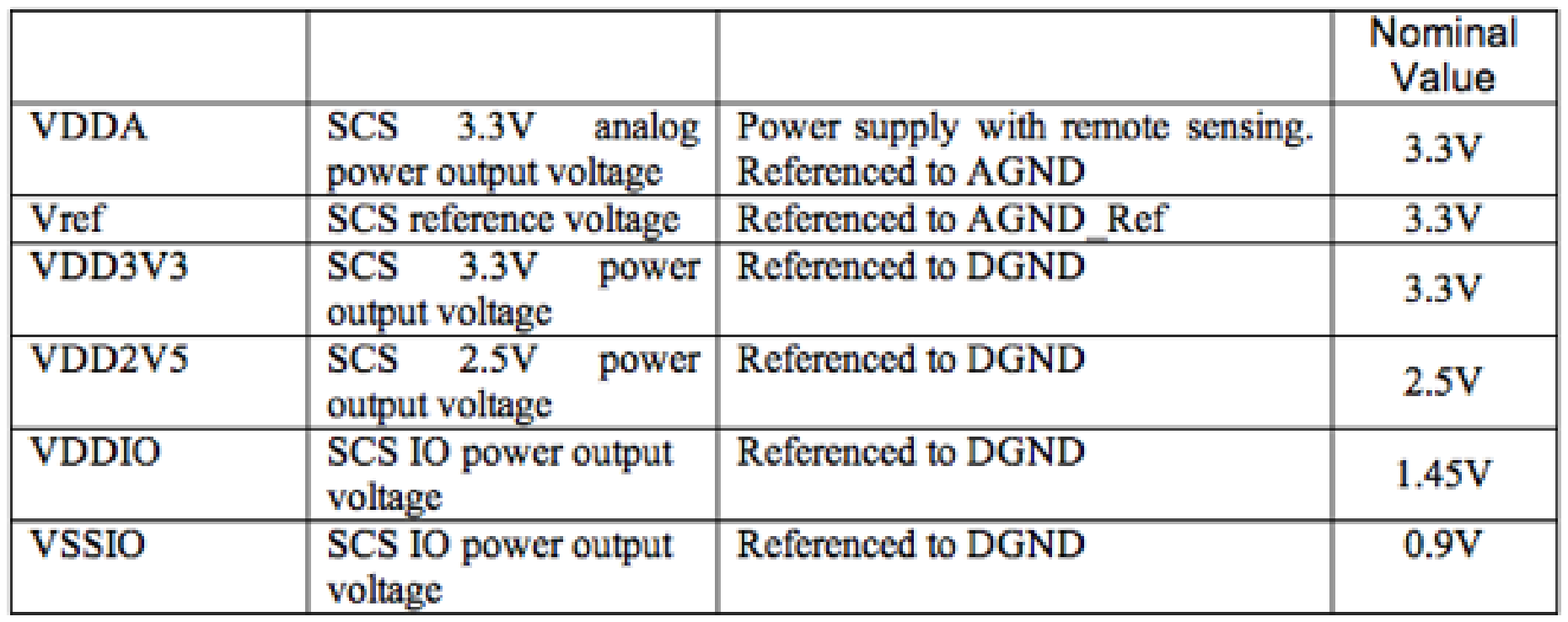}
\caption{Voltage references and grounding for proper SIDECAR ASIC feeding.}
\label{tab:4}       
\end{center}
\end{table}

\subsection{TCU unit design}
\label{TCU}
\begin{figure}[!h]
\begin{center}
 \includegraphics[width=12cm]{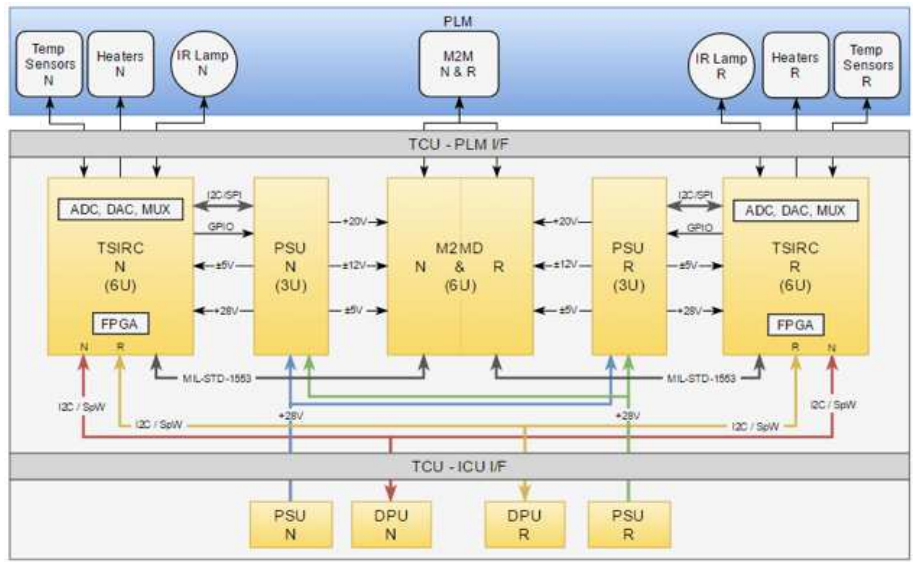}
\caption{Telescope Control Unit block diagram.}
\label{fig:7}       
\end{center}
\end{figure}

The Telescope Control Unit (refer to \cite{Sierra-Roig_1} for an exhaustive description of the metrology capabilities of the Unit) will be composed of three distinct boards as shown in Fig. ~\ref{fig:7}.

A 6U PCB (TSIRC) will hosts the PLM thermal monitoring and control HW, the IR calibration lamp driver and their multiplexing stages. For the driver electronics of M2 mechanism, it is foreseen an upgraded version of Euclid\textquotesingle s M2M, with the same driver, which will require a separated 6U board for both nominal and redundant systems (M2MD). In order to reduce M2MD modifications to fit ARIEL requirements, as well as to reduce the number of I/F from ICU\textquotesingle s PSU, simplifying it, a dedicated 3U board PSU is foreseen (TCU-PSU), which will generate (from the main power line of +28 V coming from ICU) all the voltage levels required by the M2MD and TSIRC boards. The system will be based on a cold redundancy, with all the boards resting inside a dedicated box on top of ICU\textquotesingle s, as represented in the mechanical design picture (see Fig. ~\ref{fig:9}).

The digital system of the TCU will be based on a FPGA with an embedded HDL FSM to control all the TCU boards and to simplify, at the same time, the overall SW architecture of the Unit. The UT6325 FPGA will be located in the TSIRC board as well as its PoL converters to generate the proper voltages for GPIO interfaces and internal cores. The FPGA will host two Digital Signal Processing Modules (DSPM, one for the thermal monitoring subsystem and the other one for the IR calibration lamp driver), five PID controllers, GPIO interface management to generate multiplexers addresses and select the proper voltage and gain for a given thermistor, as well as to control the OPCs of TCU-PSU. It will also include a memory bank, two I$^2$C (or SpW, option to be explored in the next B1 phase of the mission) links to communicate with DPU and one MIL-STD-1553 (or SpW as well) link to communicate with the M2M Driver.

The telescope thermal monitoring will be performed by means of two types of sensors: Cernox thermistors for precise readings (detectors, M1, optical elements, etc.) and DT-670 diodes for housekeeping TM of other elements of the PLM ($V$-grooves, OB, baffle, etc.). These sensors will be driven and read thanks to the Thermal Stabilizer \& IR Calibrator electronics. In particular, sensors will be driven by an adjustable current source (one for each type of sensor, with 4 to 6 selectable levels) and thanks to their 4-wires configuration, read by means of an instrumental amplifier (IA) and a 16-bit ADC. All 46 PLM sensors will be sequentially powered and read with a multiplexing stage with no cross-strapping: nominal sensors will be connected to nominal TSIRC and backup ones to redundant TSIRC.

Thermal perturbations in the PLM are expected to have time periods much longer than 1 s, therefore, the system will be designed so that each second all Cernox thermistors (approx. 60$\%$) plus 4$\%$ of diodes will be read so the temperature controller can be updated with the proper feedback and, every 10 s, housekeeping TM can be generated and sent to the DPU.

\begin{figure}[!h]
\begin{center}
 \includegraphics[width=12cm]{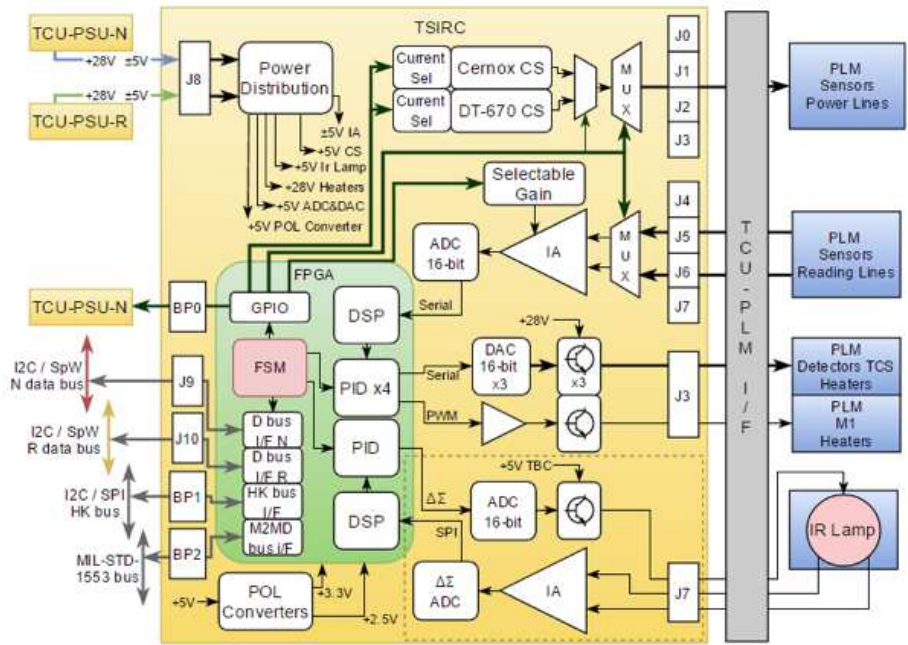}
\caption{TSIRC board block diagram.}
\label{fig:8}       
\end{center}
\end{figure}

The thermal control of the TCS subsystems (which are placed between critical detectors/mirrors and their thermal sinks) will be carried out by monitoring their temperature and activating their heaters once the correction has been calculated by the FPGA logic (a PID control type loop). The heaters power will be supplied by their driver stage, which consists of a DAC and a buffer carefully designed to supply a constant current for detector heaters (avoiding EMI with the detectors) as well as a buffered PWM signal for M1 mirror (not affected by EMI). Detectors will have a single heater to stabilize their temperature, but M1 requires several (3 to 5) to help distributing the heat. Survival heaters and thermistors might be installed in each TCS as well, but are assumed to be completely in charge of the S/C provider (their control system too). 
 
The IR calibration lamp will be based on a thermal source to generate the proper light spectrum for the detectors. Thermal source consists of a 4-wires tungsten filament in order to power and read its voltage at the same time.
ARIEL requirements foresee a 16-bits DAC to control the filament current once ground test has found the proper current to achieve 1100 K at the tungsten filament. The proposed architecture foresees a 24-bits resolution PID feedback loop in order to control the calibration lamp power with a resolution better than one part per million. Once the signal has been adapted in the instrumental amplifier, a delta sigma ADC from Texas Instruments (ADS1282-SP) will acquire a 24-bits sample every millisecond. The control logic inside the FPGA (a PID controller with a $\Delta$$\Sigma$ modulator to reduce bits count at a higher rate) will drive an overclocked ($\sim$65 MHz) 16-bits DAC, which in turn will drive a current buffer for the calibration lamp.
 
This architecture could be simplified (using 16-bits control, only reducing sampling frequency and removing the FPGA delta sigma modulator) to fulfil the ARIEL baseline requirements but could also be implemented the 24-bits control in order to gain precision when calibrating the detectors. The final solution will be chosen in the next phase of the Mission.

The M2M Driver and mechanism are based on an inherited design from Euclid\textquotesingle s and GAIA\textquotesingle s M2MM. The $\sim$4.45 kg mechanism will have 3 degrees of freedom (DOF, tip/tilt and piston) controlled by a dedicated driver hosted inside the TCU box.

The system will rely on a single 6U board within a cold redundant configuration, where nominal coils of the stepper motors are connected to the nominal section of the driver, and backup coils are connected to the redundant section. The maximum power consumption is expected to be approximately 10 W, but this value is inherited from Euclid\textquotesingle s system and might end up being higher due to the higher mass of ARIEL M2 (compared with Euclid\textquotesingle s M2). 

Drivers and mechanism voltages ($\pm$5 V, $\pm$12 V and +20 V) will be supplied by a PSU included in the TCU box. Communication with the driver board is achieved by a MIL-STD-1553 bus (or SpW, as alternative), where each section is provided with a logic capable of decoding all the telecommands received in order to generate the switching sequences required by the motors, and encoding the status information to provide serial telemetry.

\section{Harnessing}
\label{Harnessing}
Three different sets of harnesses are foreseen: internal to the ICU and TCU subsystems (internal harnessing), towards the Payload and towards the S/C (or external harnessing for the SVM-Payload linking).
The SIDECAR-DCU electrical I/F as well as the adopted kind of harnessing has been already described in the previous paragraphs; hereunder are explicated the ICU internal I/F as well as those towards the S/C.

\subsection{Internal harnessing}
The main electrical I/F internally connecting the ICU and TCU subsystems are the following:

\begin{itemize}
\renewcommand{\labelitemi}{\tiny$\bullet$}
\item	DPU Power I/F:	 +5 V		(from PSU to DPU)
\item	DCU Power I/F:	 +5 V		(from PSU to DCU$_0$ and DCU$_1$)
\item	TCU Power I/F: +28 V	(from PSU to TCU N and R PSU boards)

\end{itemize}

\begin{itemize}
\renewcommand{\labelitemi}{\tiny$\bullet$}
\item	DPU TM/TC I/F:	GPIO and SPI		(from DPU to PSU)
\item	DCU TM/TC I/F:	RS485, SPI and SpW	(from DPU to DCU$_0$ and DCU$_1$)
\item	TCU TM/TC I/F:	I$^2$C, SPI or SpW	(from DPU to TCU N and R TSIRC boards)

\end{itemize}

In case a back panel were adopted for signal and power lines routing inside the ICU and/or TCU boxes (baseline choice) no flying harness would be foreseen between their boards. Note that the ICU-TCU harnessing will be implemented by means of external (i.e. outside the ICU and TCU boxes) connections. This choice would facilitate the AIV and AIT activities at Unit and System level, as already pointed out.

\subsection{External harnessing}
Concerning the selected (electrically and thermally) conductive materials, the harnessing towards the Payload (AIRS, telescope mirrors and mechanisms) is partially to be refined as it plays a fundamental role in thermal linking the warm electronics side to the cold electronics part (it shall be further assessed during the next Phase). The foreseen harnessing towards the S/C is hereunder itemized.\\

Nominal (and Redundant) Power Supply and control I/F:
\begin{itemize}
\renewcommand{\labelitemi}{\tiny$\bullet$}
\item	Power line: +28 V + RTN			(From S/C PCDU to ICU PSU)
\item	Switch On HPC (Signal + RTN)		(From S/C DMS to ICU PSU)
\item	Switch Off HPC (Signal + RTN)		(From S/C DMS to ICU PSU)
\item	Switch Status BSM\footnote{Bi-level Switch Monitor} (Signal + RTN)		(From S/C DMS to ICU PSU)
\item	Sync (Signal + RTN)				(From S/C DMS to ICU DPU)

\end{itemize}

Nominal (and Redundant) I/O digital TM/TC I/F:
\begin{itemize}
\renewcommand{\labelitemi}{\tiny$\bullet$}
\item	1 or 2 (baseline 1 TM + 1 TC) Standard SpaceWire links (configured @ 10 Mbit/s) (From S/C DMS to the ICU DPU).
\end{itemize}

The MIL-STD-1553 bus use is not foreseen at this stage of the design, except for the M2M driver TM/TC towards the TSIRC board.

\section{Mechanical design}
\label{Mech}
\begin{figure}[!h]
\begin{center}
 \includegraphics[width=10cm]{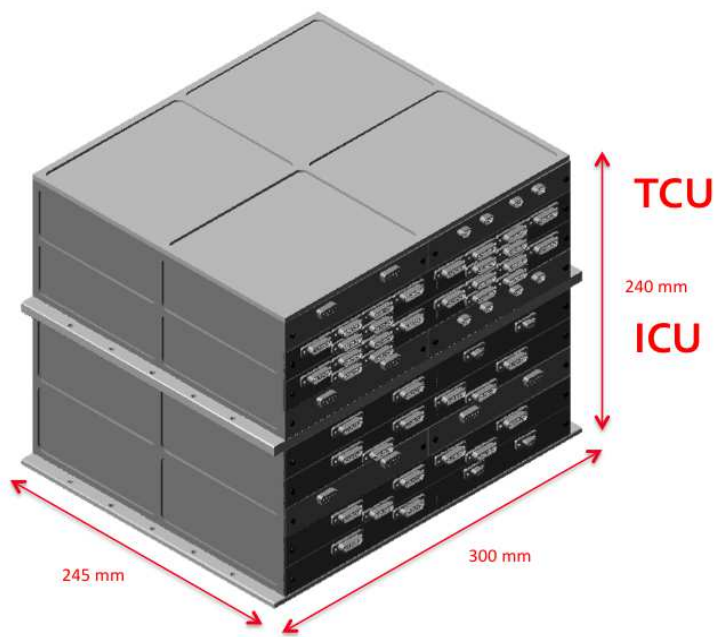}
\caption{ICU and TCU boxes baseline mechanical design in a stacked configuration.}
\label{fig:9}       
\end{center}
\end{figure}

Fig. ~\ref{fig:9} shows the 3D CAD model of the overall Unit (ICU and TCU), which foresees two stacked boxes hosting:

\begin{itemize}
\renewcommand{\labelitemi}{\tiny$\bullet$}
\item	2x ICU/PSU (N\&R), in 3U format
\item	2x DPU (N\&R), in 6U format
\item	2x DCU (both N), in 6U format

\end{itemize}

inside the ICU, and:

\begin{itemize}
\renewcommand{\labelitemi}{\tiny$\bullet$}
\item	2x TSIRC (N\&R), in 6U format
\item	2x M2M (N\&R) drivers, in 3U format (TBC)
\item	2x TCU/PSU (N\&R), in 3U format
\end{itemize}

inside TCU. 

Indeed, the M2M control logic N\&R could be implemented in a single 6U-format PCB without any change on the TCU mechanical envelope.

The Unit overall dimensions are 300 mm (including mounting feet) x 240 mm x 245 mm. The boxes depth (245 mm) also foresees the adoption of two distinct back panels for ICU and TCU power and signals lines routing.

Both boxes shall host the grounding reference point along with a bounding stud and at least a TRP (Temperature Reference Point). Their mechanical coupling will be done joining the bottom box top plate with the top box base plate, assuring a proper thermal conduction and heat dissipation to the SVM optical bench by means of the lateral panels.

Connectors shown in the mechanical design are only indicative (MDM micro-d type, 9 and 25 poles, for SpW TM/TC and analog/digital signals; DSUB, 9 poles, for power and control signals).

\section{ICU alternative architecture}
\label{ICU_alternative}
The ICU alternative solution is very similar to the baseline one and is designed for the AIRS Spectrometer with the US detectors or European detectors ($CH_0$ and $CH_1$ channels) mainly thanks to different DCU control electronics.

EU sensors are developed within a joint collaboration between CEA-LETI and Sofradir (ROIC development) and characterized by two different pixel architectures and readout modes under the AIRS Team responsibility.

In this configuration, the AIRS architectural blocks of the electronics design are the following:

\begin{itemize}
\renewcommand{\labelitemi}{\tiny$\bullet$}
\item Detector Sensor Chip Assembly + Read Out Integrated Circuit (AIRS FPA);
\item	Cold Front End Electronics (Payload side);
\item	Warm Front End Electronics (DCU on SVM side);
\item	Instrument Control Unit (and TCU unit on SVM side).

\end{itemize}

\begin{figure}[!h]
\begin{center}
 \includegraphics[width=12cm]{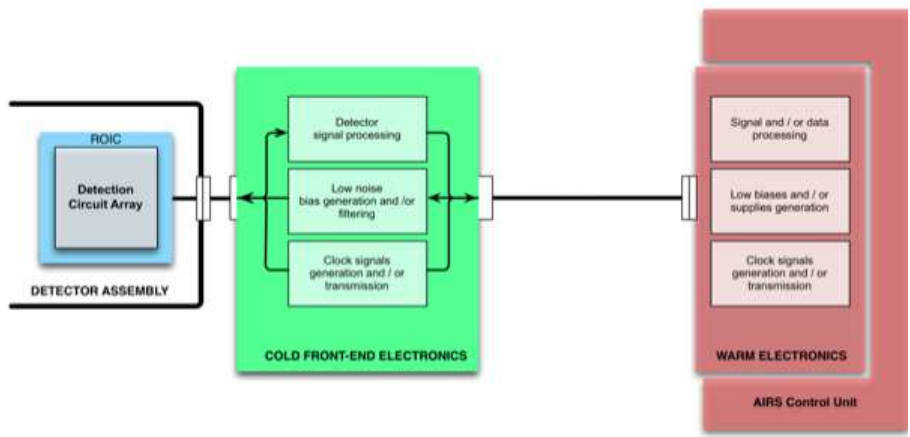}
\caption{AIRS detectors-CFEEs-WFEEs alternative solution block diagram.}
\label{fig:10}       
\end{center}
\end{figure}

The adoption of European detectors and ASICs/CFEEs is pending on TRL level that shall reach level 6 before freezing the adopted technology for the electronics chain design.

The two pixel alternatives for the European detectors are SFD (Source Follower per Detector) and CTIA (Capacitance Trans Impedance Amplifier), respectively requiring the following CFEEs (presently in charge of SRON):

\begin{itemize}
\renewcommand{\labelitemi}{\tiny$\bullet$}
\item	In the SFD case an ASIC would be required;
\item	In the CTIA case the baseline choice would be to adopt an amplification and A/D conversion stage located on an intermediate 110 K thermal interface.

\end{itemize}

The preferred option relies on the adoption of a CTIA-based pixel readout architecture plus a 110 K readout electronics stage (with performance expected to be in the 30 to 50 $e^-$ $rms$ as readout noise) for the European detectors CFEE, as shown in Fig. ~\ref{fig:10}.
In the first case, ROIC is SFD type and requires a SIDECAR CFEE.

\begin{figure}[!h]
\begin{center}
 \includegraphics[width=12cm]{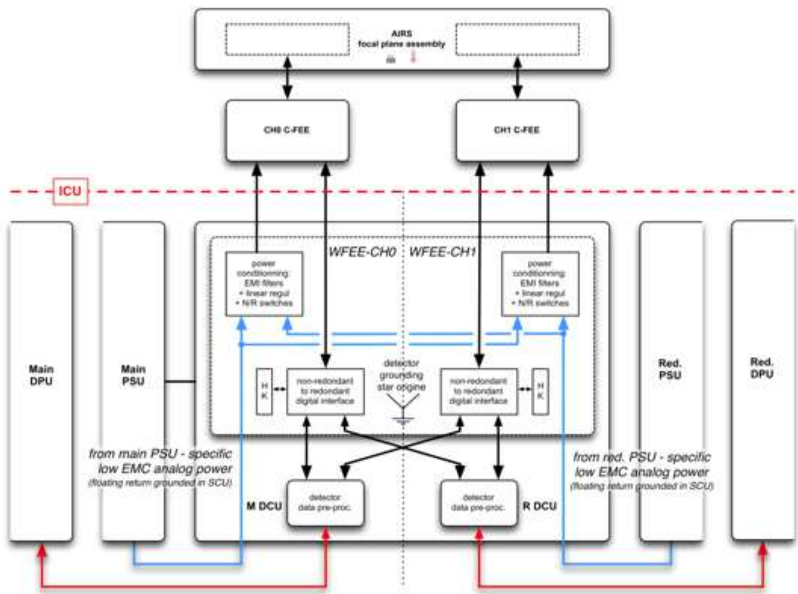}
\caption{ICU alternative solution showing details concerning WFEE redundancy and cross-strapping.}
\label{fig:11}       
\end{center}
\end{figure}

The ICU alternative design block diagram is represented in Fig. ~\ref{fig:11}. The DCU Unit is depicted by means of a separate block (which can be internal or external to the ICU box) hosting the digital I/F to the Spectrometer cold front end electronics (ASIC or CFEE stage) and the power conditioning sections for both $CH_0$ and $CH_1$.

TCU and PSU are respectively implemented with 2 x 3U + 1 x 6U and 1 x 6U format boards.

\subsection{DCU alternative design}
\label{DCU_alternative}
The DCU can be in the form of integrated boards into the ICU (preferred option) or in a distinct mechanical box or drawer (that in principle could be stacked to the ICU and TCU boxes) and will host the following functionalities:

\begin{itemize}
\renewcommand{\labelitemi}{\tiny$\bullet$}
\item	CFEE/ASIC control and configuration;
\item	Detector readout and control sequencing;
\item	Digital data processing and packaging;
\item	Low-level command decoding;
\item	Low-level command ACK and HK parameters \& data packets transfer;
\item	Clean power generation for CFEEs;
\item	PSU interfacing along with the needed cross-strapping.

\end{itemize}

\begin{figure}[!h]
\begin{center}
 \includegraphics[width=12cm]{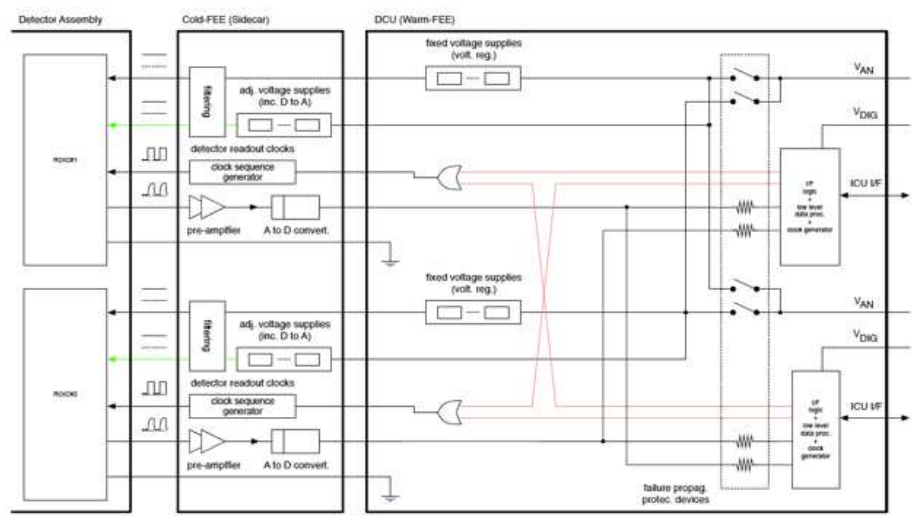}
\caption{DCU internal architecture (US/Teledyne detectors and SIDECAR case).}
\label{fig:12}       
\end{center}
\end{figure}

\begin{figure}[!h]
\begin{center}
 \includegraphics[width=12cm]{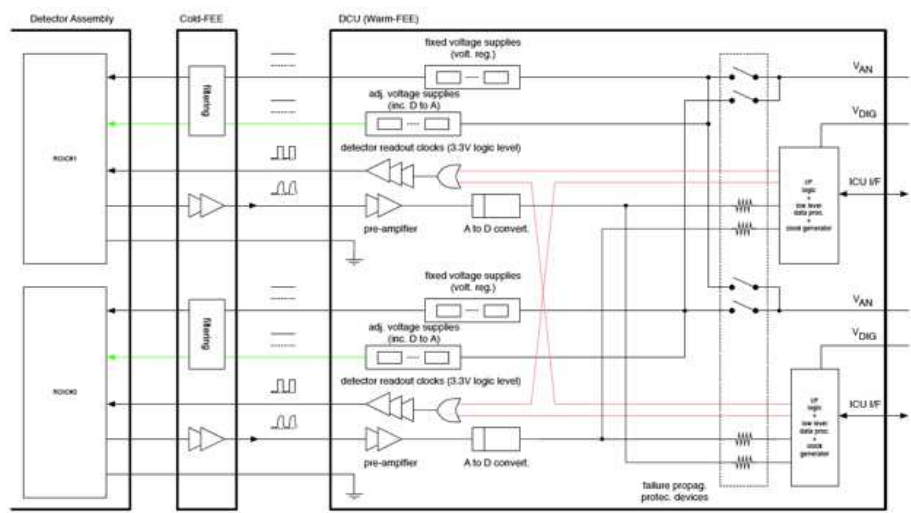}
\caption{DCU internal architecture (European detectors and CFEE case).}
\label{fig:13}       
\end{center}
\end{figure}

Fig. ~\ref{fig:12} and Fig. ~\ref{fig:13} show two viable architectures in case of adoption of US or European detectors and CFEE.

The DCU alternative electrical design is still to be refined but could implement a processor\footnote{In order to implement only a high-level SW (ASW) running on the ICU processor (hosted by DPU) and in charge of Instrument management, data processing (TBD) and FDIR procedures, this complex solution should be avoided.} and/or a space qualified FPGA. As for the ICU internal boards, a PoL section is needed to derive the required voltage levels for the on-board devices and electronics components.

Due to the use of the Teledyne SIDECAR ASIC in the cold front-end electronics that already implements many functions, the DCU has a low level of complexity.
 Beside the functions that interface the non-redundant detector assemblies to the rest of the electronics, the warm front-end electronics comprises redundant FPGA that implements low level SIDECAR commands generation and scientific data buffering. DCU interfaces externally with the DPU function of the ICU through a SpW serial interface.

In addition to the functions already mentioned above, the DCU for the EU detectors solution implements a clock sequencer function that provides both analog electronics and detector with signals that sequence the detection chain/assembly clocks, since AIRS-CFEE does not feature digital to analog converters so, these devices, shall be located within DCU.

\section{ICU Electrical Ground Support Equipment }
\label{EGSE}
The short-functional, full-functional and performance tests on the overall ICU assembly shall be performed using an appropriate Electrical Ground Support Equipment (EGSE). The EGSE shall support both the ICU test/verification and the AIRS Spectrometer end-to-end test at different stages of the AIV flow, e.g.:

\begin{itemize}
\renewcommand{\labelitemi}{\tiny$\bullet$}
\item	ICU integration and testing (using additional test equipment to simulate the instrument I/F); 
\item	Integration and testing of the WFEE (DCU) + CFEE + FPA (AIRS-level); 
\item	Overall ARIEL payload Integration and Verification (TCU functionalities and telescope monitoring included).

\end{itemize}

At the first stage, the EGSE will include additional HW and SW Test Equipments (TE) simulating the relevant I/F and functionalities of the missing payload segments. These additional TE will be totally or partyially removed from the EGSE as soon as the related units will be added to the test configuration.

The ICU subsystems simulators and EGSE are needed to perform the preliminary tests (Short Functional Tests, Full Functional Tests) on the ICU Engineering Model (EM) and later on the FM/PFM Model (Performance Tests), as the chosen baseline model philosophy for the ICU subsystem is the Proto-Flight approach.

\subsection{Overview}
\label{ovw} 
\begin{figure}[!h]
\begin{center}
 \includegraphics[width=8cm]{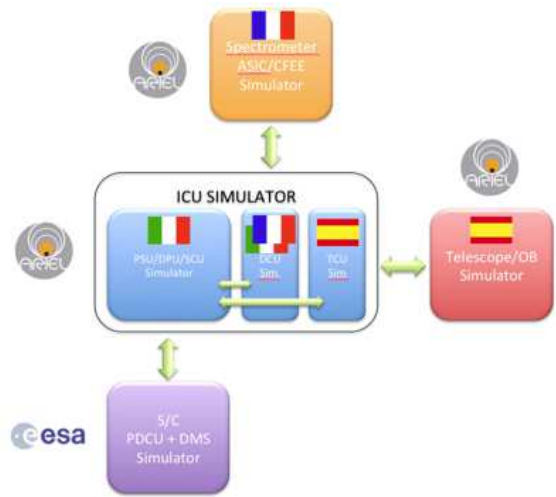}
\caption{ARIEL ICU needed simulators and EGSE; green arrows represent the electrical I/F.}
\label{fig:14}       
\end{center}
\end{figure}

The aim of the ICU EGSE is to support testing and operations on both the ICU and the AIRS Spectrometer. A block diagram of the needed ICU subsystems simulators and EGSE is illustrated in Fig. ~\ref{fig:14}. The represented scheme fits both the baseline and the alternative solution design.

The main functions performed by the ICU EGSE may be summarized as follows:

\begin{itemize}
\renewcommand{\labelitemi}{\tiny$\bullet$}
\item TM/TC S/C interface exercising; 
\item	Storage of scientific data (S/C science data interface to mass memory simulator); 
\item	Power generation; 
\item	Payload instrument simulation with possibility to simulate instrument HK data; 
\item	ROICs/ASICs/CFEEs simulation with possibility to simulate Scientific data; 
\item	Data packets acquisition for storage and following monitoring \& processing. 

\end{itemize}

The foreseen GSE (Electrical and SW GSE) is the following: 

\begin{itemize}
\renewcommand{\labelitemi}{\tiny$\bullet$}
\item	ICU EGSE: 
\subitem	- Workstations (desktops and/or laptops PC);
\subitem	- SCOS-2000/SCOE (ESA provided);
\subitem	- Harnessing (to the S/C and to the Payload);
\subitem	- Software (S/C and ICU subsystems simulators);
\subitem	- Instrument Database (IDB hosting TCs, TMs and related parameters);
\subitem	- Instrument Databank/Datapool;
\subitem	- S/C Interfaces Simulator (SIS).

\end{itemize}

\subsection{Functional description}
\label{func_desc} 
The ICU EGSE shall allow the test operator or Test Conductor to:

\begin{itemize}
\renewcommand{\labelitemi}{\tiny$\bullet$}
\item	Fully control the ICU via a unique control station; 
\item	Command \& monitor the ICU via the adopted S/C communication protocol (SpW / CCSDS / RMAP) through a DMS simulator (SIS); 
\item	Generate primary power bus and interface all HK lines; 
\item	Send simulated scientific data via ROICs / ASICs / CFEEs Data \& HK simulators; 
\item	Acquire the ICU scientific packets via the Data Acquisition link (SSMM / OBC or DMS simulator); 
\item	Compare the acquired data versus the simulated data to check for ICU correct data handling.

\end{itemize}

The best candidate for the SW platform to be adopted for the EGSE is the standard ESA SCOS-2000 (Spacecraft Control \& Operations System). SCOS-2000 is the generic mission control system software of ESA, which supports CCSDS TM and TC packet standards and the ESA Packet Utilization Standard (PUS). It has been proven by recent ESA missions (e.g. Herschel/Planck) that, using SCOS-2000 and its add-ons, it can be extended to cover the on-ground testing phase and work as a proper EGSE.

The use of SCOS-2000 will guarantee a very smooth transition from the ICU subsystem AIV to the ARIEL Payload AIV and the in-orbit operation phases (assuming that SCOS-2000 will be adopted by ESA for the EGSE and the Ground Segment, respectively). In particular, this smooth transition will concern the SCOS-2000 instrument Data Base (MIB tables), which describes the TM and TC packets structure.

\section{On-Board Software }
\label{OBSW}
The ARIEL ICU On Board Software (OBSW) \cite{Farina_1} will be composed of the following three main components:

\begin{enumerate}
\item Basic Software: 

\begin{itemize}
\renewcommand{\labelitemi}{\tiny$\bullet$}
\item	\emph{Boot software}: it is installed on the PROMs of the ICU DPU board and allows loading the ICU Application Software. It contains all the low-level drivers for the CPU board and its related interfaces. 
\item	\emph{Basic I/O SW, Service SW \& Peripheral Drivers}: it is a HW-dependent Software including the Software Drivers for all the internal and external ICU digital interfaces. This SW is used by the Application SW and can depend on the selected Operating System (OS).
\end{itemize}

\item Application Software:
\begin{itemize}
\renewcommand{\labelitemi}{\tiny$\bullet$}
\item \emph{Instrument Control \& Configuration Software}: it implements the ARIEL scientific payload handling. It controls the spectrometer, implements the operating modes, monitors the instrument health and runs FDIR procedures. It implements the interface layer between the S/C and the instrument. 
\item	\emph{Data Processing and Compression Software}: it implements all the necessary on-board processing functionalities, included the on-board lossless compression (if needed). After the processing the SW prepares CCSDS packets for the transmission to the S/C Mass Memory (SSMM).
\end{itemize}

\item Real-Time Operating System (RTOS): the selected baseline operating system is RTEMS. 

\end{enumerate}

\begin{figure}[!h]
\begin{center}
 \includegraphics[width=10cm]{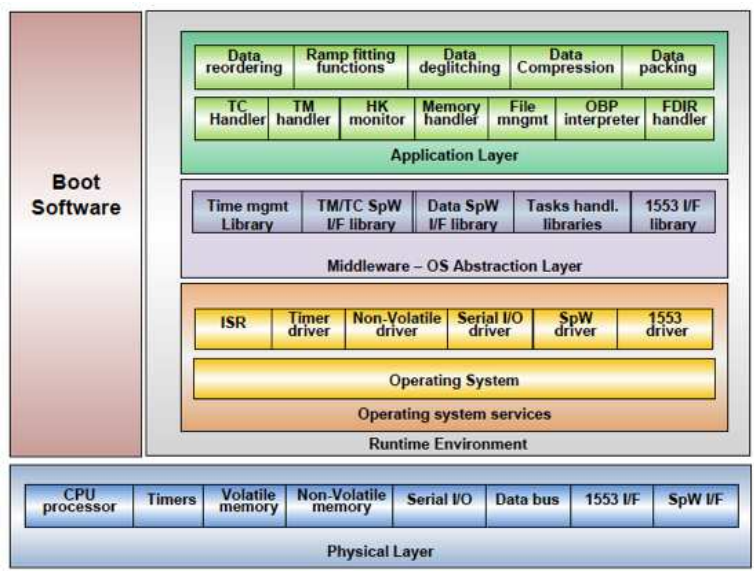}
\caption{SW layers structure of the ARIEL On Board Software.}
\label{fig:15}       
\end{center}
\end{figure}

The role and the interconnections between the three listed components can be clearly identified in the layered representation reported in Fig. ~\ref{fig:15}.

\subsection{On-board software layers }
\label{SW_layers} 
With reference to the following block diagram, the physical layer includes all the ICU HW components with a direct level of interaction with the on-board software. The Runtime environment includes the Real Time Operating System layer, necessary to provide multi-tasking support. In case the baseline architecture based on the LEON processor will be confirmed, the RTEMS operating system is a good RTOS candidate, being already used for applications on board ESA satellites. The other indicated system services are those not directly provided with the OS kernel, but included in the Basic Software Component mentioned above.

An OS abstraction layer has then been included in the layered structure for the ARIEL OBSW, in which all middleware libraries have been considered. The middleware services are based on the use of RTOS function calls. They include all library functions dedicated to the low-level handling of the ICU HW devices/interfaces. All the middleware libraries will be developed in house and will provide a mean for developing the Application Software virtually independent from the HW and OS below it. This layer is very important and will ease the testing activities.

The Application Layer includes both the ICU Instrument Control software and the Data Processing software.

The ICU \emph{Instrument Control SW} will implement the TM/TC S/C interface handling, the payload housekeeping data acquisition and monitoring, the instruments operating modes management and the autonomous function execution. The software will be written in $C$, though some functions may need to be coded in assembly to optimize their performance.

In case stringent timing requirements have to be met for subsystem commanding, an interrupt driven command sequencer (On Board procedures, OBP interpreter) can be included into the ICU on board software. Based on the experience of Herschel\textquotesingle s HIFI and SPIRE instrument control software, this is a flexible and effective solution to implement time-critical commanding procedures.

The \emph{Data Processing SW} implements all the necessary on board processing functionalities, included (if implemented) the on-board lossless compression (i.e. RICE). After the processing the SW prepares CCSDS packets for the transmission to the S/C Solid State Mass Memory.

For the ARIEL Science it is desirable to minimize the on-board data processing. To allow the Science Team to have the optimum chance to extract the best SNR from the available data, the capability to improve the processing during the mission and the maximum flexibility in the algorithms can be exploited using more complex on-ground processing.

The actual processing will be finalized during the Phase B study exploiting simulated data flows to verify the effectiveness of the adopted data reduction steps for the selected detectors. In particular, the deglitching algorithm performances shall be verified against the expected data redundancy (spectra overlapping) as well as the data acquisition rate and the $spaxels$\footnote{A $spaxel$ is a set of binned pixels in both spatial and spectral dimensions.} dimensions.

Finally, the need to implement an on-board effective lossless compression is strictly related to the results of the on-board deglitching algorithm. If required, a dedicated trade off activity to evaluate the performances of different standard lossless compression algorithms on the on-board CPU processor shall be planned.

\section{Conclusions}
\label{Conc}
The presented ICU baseline electrical architecture can be considered as the best solution for what concern the risks assessment and mitigation during the present Phase, as the very high TRL of its subsystems as well as the computed reliability figure for the DPU / DCU / SIDECAR / Detector chain, larger than 98$\%$, lead to an overall design characterized by already developed and tested boards.

The strong heritage of the Teledyne detectors and SIDECAR ASIC, coupled with the lessons learned from the development and testing of the DCU boards up to the EQM model and their electrical I/F for the Euclid Mission, will guarantee a very reliable Unit based on already available on the shelf subsystems.
On the contrary, an early management of the ITAR procedures for the required US subsystems shall be undertaken.

Finally, the power, mass and volume budges derived for the baseline solution, as well as its complexity, are compliant with those allocated by ESA for the Unit, margins included.

Concerning the alternative design, its main advantage is to have the full contributors to the AIRS performance under the same system responsibility (DCU requirements specification by the AIRS Team) ensuring that the design is optimal in terms of adaptation of the solution to the needs of the Mission. It also ensures that the full data acquisition chain can be verified before delivery.

The alternative architecture also ensures that, independently of the detector choice at the end of phase B1, the external interfaces to ICU remain identical, although it should be noted that the ICU baseline architecture, with its own flexibility granted by the DCU design, would allow driving ASICs and detectors different from the US ones, likely without any modification or only with marginal changes.

On the other hand, the alternative electronics design could be more complex especially for what concerns the electrical interfaces, as a further box for DCUs could be required between ICU and AIRS.

Depending on the final DCU implementation (in a separate box/drawer or as an internal ICU board) the mass, volume and power budgets could be impacted and might require a further revision.

\begin{acknowledgements}

The authors gratefully acknowledge the Italian Space Agency for the financial contribution to the ARIEL project in the framework of the ASI-INAF agreement 2015-038-R.0 and for the financial support from the Spanish Ministry of Economy and Competitiveness (MINECO) through grants ESP2014-57495C2-2-R and ESP2016-80435-C2-1-R.

A special thank to the European Space Agency for the support provided by the ARIEL Study Team, to the University College of London (UCL) leading the Project and to the Rutherford Appleton Laboratory (RAL Space) Managers and Engineers.

\end{acknowledgements}


\begin{thebibliography}{}
%
%

\bibitem{Tinetti_1}
G. Tinetti et al, \emph{}, Special Issue on ARIEL, Experimental Astronomy, (2017)

\bibitem{Eccleston_1}
P. Eccleston et al, \emph{}, Special Issue on ARIEL, Experimental Astronomy, (2017)

\bibitem{Morgante_1}
G. Morgante et al, \emph{}, Special Issue on ARIEL, Experimental Astronomy, (2017)

\bibitem{Da_Deppo_0}
V. Da Deppo et al, \emph{}, Special Issue on ARIEL, Experimental Astronomy, (2017)

\bibitem{Amiaux_1}
J. Amiaux et al, \emph{}, Special Issue on ARIEL, Experimental Astronomy, (2017)

\bibitem{Da_Deppo_1}
V. Da Deppo et al, \emph{An afocal telescope configuration for the ESA ARIEL mission}, ICSO International Conference on Space Optics, Biarritz (FR), (2016)

\bibitem{Da_Deppo_2}
V. Da Deppo et al, \emph{The afocal telescope optical design and tolerance analysis for
the ESA ARIEL Mission}, OSA Optical Design and Fabrication Congress, Denver (US), (2017)

\bibitem{Da_Deppo_3}
V. Da Deppo et al, \emph{An afocal telescope configuration for the ESA ARIEL Mission
}, CEAS Aeronautical Journal, (2017)

\bibitem{Rataj_1}
M. Rataj et al, \emph{}, Special Issue on ARIEL, Experimental Astronomy, (2017)

\bibitem{Sierra-Roig_1}
C. Sierra-Roig et al, \emph{The ARIEL ESA mission on-board metrology}, Proceedings of the 4\textsuperscript{th} IEEE International Workshop on Metrology for Aerospace, Padova (IT), (2017)

\bibitem{Farina_1}
M. Farina et al, \emph{Ariel Spectrometer Instrument Control and Data Processing Software}, European Planetary Science Congress (EPSC), Riga (LT), (2017)


\end{thebibliography}


\end{document}